\documentclass{article}
\usepackage{amssymb}
\usepackage{amsmath}
\usepackage[dvips]{graphicx}

\input{tcilatex}

\begin{document}

\title{Charged Fermions Tunnelling from Kerr-Newman Black Holes}
\author{Ryan Kerner\thanks{%
rkerner@uwaterloo.ca} and R.B. Mann\thanks{%
rbmann@sciborg.uwaterloo.ca} \\
Department of Physics \&\ Astronomy, University of Waterloo\\
Waterloo, Ontario N2L 3G1, Canada}
\maketitle

\begin{abstract}
We consider the tunnelling of charged spin-$\frac{1}{2}$ fermions from a
Kerr-Newman black hole and demonstrate that the expected Hawking temperature
is recovered. We discuss certain technical subtleties related to the
obtention of this result.

\begin{description}
\item[PACS:] 04.62.+v, 04.70.Dy,

\item[Keywords:] black holes, Kerr-Newman, fermions, tunnelling, Hawking
temperature
\end{description}
\end{abstract}

\section{Introduction}

Semi-classical methods of modeling Hawking radiation as a tunnelling effect
were developed over the past decade and have garnered a lot of interest \cite%
{early tunnelling}-\cite{ferm kerr}. \ The earliest work with black hole
tunnelling was done by Kraus and Wilczek \cite{early tunnelling}, and was
refined by various researchers \cite{late 90s tunnelling,Parikh, Padmanabhan}%
. From this approach an alternative way of understanding black hole
radiation emerged. In particular\textbf{\ }one can calculate the Hawking
temperature in a manner independent of traditional Wick Rotation methods\ or
the original method of modelling gravitational collapse \cite{hawking}. \
Tunnelling provides not only a useful verification of thermodynamic
properties of black holes but also an alternate conceptual means for
understanding the underlying physical process of black hole radiation. \ For
scalar field emission it has been shown to be very robust, having been
successfully applied to a wide variety of interesting and exotic spacetimes,
including the Kerr and Kerr-Newman cases \cite{kerr and kerr newman, Zhang
and Zhao, first paper}, black rings \cite{Black Rings}, the 3-dimensional
BTZ black hole \cite{Vanzo, BTZ}, the Vaidya spacetime \cite{Vaidya}, other
dynamical black holes \cite{dynamicalbh}, Taub-NUT spacetimes \cite{first
paper}, G\"{o}del spacetimes \cite{Godel}, and dynamical horizons \cite%
{dynamicalbh}. \ Tunnelling methods have even been applied to horizons that
are not black hole horizons including those with cosmological horizons \cite%
{parikh1},\cite{schwdS},\cite{ferm cosmo hor},and Rindler Spacetimes \cite%
{Padmanabhan},\cite{first paper},\cite{Fermion tunnelling} for which it has
been shown the Unruh temperature \cite{Unruh} is in fact recovered.

In general tunnelling methods\ involve calculating the imaginary part of the
action for the (classically forbidden) process of\ s-wave emission across
the horizon, which in turn is related to the Boltzmann factor for emission
at the Hawking temperature. \ Different approaches exist for calculating the
imaginary part of the action for the emitted particle. \ The first black
hole tunnelling method developed was the Null Geodesic Method used by Parikh
and Wilczek \cite{Parikh}, which followed from the work of Kraus and Wilczek 
\cite{early tunnelling}. \ The other approach to black hole tunnelling is
the Hamilton-Jacobi Ansatz used by Agheben et al \cite{Vanzo}, which is an
extension of the complex path analysis of Padmanabhan et al \cite%
{Padmanabhan}. \ Recently we have extended black hole tunnelling to include
the emission of spin 1/2 fermions \cite{Fermion tunnelling}, demonstrating
that fermion tunnelling from both a Rindler horizon and a generic
non-rotating black hole recovers the expected results for temperature. While
perhaps not surprising, the result is non-trivial insofar as fermionic vacua
are distinct from bosonic vacua, and can lead to distinct physical results 
\cite{alsingpaper}. \ This work has been extended to describing fermionic
tunnelling across horizons in other spacetimes such as\ the BTZ black hole 
\cite{ferm BTZ}, and dynamical horizons \cite{ferm variable mass}. \ 

All tunnelling approaches use the fact that the WKB approximation of the
tunnelling probability for the classically forbidden trajectory from inside
to outside the horizon is given by:%
\begin{equation}
\Gamma \propto \exp (-2\,\mathrm{Im}I)  \label{Tunprob}
\end{equation}%
where $I$ is the classical action of the trajectory to leading order in $%
\hslash $\ (here set equal to unity). \ Where these methods differ is in how
the action is calculated. For the Null Geodesic method the only part of the
action that contributes an imaginary term is $\int_{r_{in}}^{r_{out}}p_{r}dr$%
, where $p_{r}$is the momentum of the emitted null s-wave. \ Then by using
Hamilton's equation and knowledge of the null geodesics it is possible to
calculate the imaginary part of the action. \ For the Hamilton-Jacobi ansatz
it is assumed that the action of the emitted (scalar) particle satisfies the
relativistic Hamilton-Jacobi equation. From the symmetries of the metric one
picks an appropriate ansatz for the form of the action and plugs it into the
Relativistic Hamilton-Jacobi Equation to solve. (For a detailed comparison
of the Hamilton-Jacobi Ansatz and Null-Geodesic methods see \cite{first
paper}). \ The Hamilton-Jacobi Ansatz came from applying the WKB
approximation to the Klein-Gordon equation. \ To lowest order in WKB this
results in the Hamilton-Jacobi equation. \ For tunnelling of spin 1/2
particles it can be shown that applying the WKB approximation to the Dirac
equation\ instead of the Klein-Gordon equation yields the tunnelling
probability for fermions \cite{Fermion tunnelling}. \ This was the first
time that the tunnelling approach had been used to model spin 1/2 particles.
\ 

In this paper we extend the tunnelling method to model charged spin 1/2
particle emission from rotating black holes. To this end we apply the
fermion tunnelling method to the Kerr-Newman black hole for both massless
and massive charged particle emission. This extension introduces some
non-trivial technical features associated with the choice of $\gamma $
matrices. We confirm that spin 1/2 fermions are emitted at the expected
Hawking Temperature from rotating black holes, providing further evidence
for the universality of black hole radiation.

\section{Charged Spin 1/2 Particle Emission From Kerr-Newman Black Holes}

We will consider particle emission from the Kerr-Newman solution. \ The
Kerr-Newman metric and vector potential are given by%
\begin{align}
ds^{2}& =-f(r,\theta )dt^{2}+\frac{dr^{2}}{g(r,\theta )}-2H(r,\theta
)dtd\phi +K(r,\theta )d\phi ^{2}+\Sigma (r,\theta )d\theta ^{2}  \notag \\
A_{a}& =-\frac{er}{\Sigma (r)}[(dt)_{a}-a\sin ^{2}\theta (d\phi )_{a}] \\
f(r,\theta )& =\frac{\Delta (r)-a^{2}\sin ^{2}\theta }{\Sigma (r,\theta )}, 
\notag \\
g(r,\theta )& =\frac{\Delta (r)}{\Sigma (r,\theta )},  \notag \\
H(r,\theta )& =\frac{a\sin ^{2}\theta (r^{2}+a^{2}-\Delta (r))}{\Sigma
(r,\theta )}  \notag \\
K(r,\theta )& =\frac{(r^{2}+a^{2})^{2}-\Delta (r)a^{2}\sin ^{2}\theta }{%
\Sigma (r,\theta )}\sin ^{2}(\theta )  \notag \\
\Sigma (r,\theta )& =r^{2}+a^{2}\cos ^{2}\theta  \notag \\
\Delta (r)& =r^{2}+a^{2}+e^{2}-2Mr  \notag
\end{align}

Since the tunnelling method is not applicable to extremal black holes \cite%
{first paper}, we will assume a non-extremal black hole so that $%
M^{2}>a^{2}+e^{2}$. Consequently there are two horizons located at $r_{\pm
}=M\pm \sqrt{M^{2}-a^{2}-e^{2}}$. \ It is convenient for our calculations to
work with the function $F(r,\theta )=-\left( g^{tt}\right) ^{-1}$ where

\begin{equation}
F(r,\theta )=f(r,\theta )+\frac{H^{2}(r,\theta )}{K(r,\theta )}=\frac{\Delta
(r)\Sigma (r,\theta )}{(r^{2}+a^{2})^{2}-\Delta (r)a^{2}\sin ^{2}\theta }
\label{e2}
\end{equation}%
and where the angular velocity at the black hole horizon is

\begin{equation}
\Omega _{H}=\frac{H(r_{+},\theta )}{K(r_{+},\theta )}=\frac{a}{%
r_{+}^{2}+a^{2}}  \label{e3}
\end{equation}

We will only show the calculation explicitly for the spin up case; the final
result is also the same for the spin down case as can be easily shown using
the methods described below. \ In the non-rotating case a statistical
argument was used to justify the assumption that overall a zero angular
momentum state is maintained for fermion emission, because as many particles
with spin pointing radially outward (spin up) would be emitted as particles
with spin pointed radially inward (spin down). \ This argument is still
valid in the rotating case: the statistical distribution of spins in the
fermion emission spectrum should not alter the angular momentum of the black
hole.\ 

The Dirac equation with electric charge is:

\begin{equation}
i\gamma ^{\mu }(D_{\mu }-\frac{iq}{\hbar }A_{\mu })\psi +\frac{m}{\hbar }%
\psi =0  \label{Dirac}
\end{equation}%
where: 
\begin{eqnarray}
D_{\mu } &=&\partial _{\mu }+\Omega _{\mu }  \label{D1} \\
\Omega _{\mu } &=&\frac{1}{2}i\Gamma _{\text{ \ }\mu }^{\alpha \text{ \ }%
\beta }\Sigma _{\alpha \beta }  \label{D2} \\
\Sigma _{\alpha \beta } &=&\frac{1}{4}i[\gamma ^{\alpha },\gamma ^{\beta }]
\label{D3}
\end{eqnarray}

The $\gamma ^{\mu }$ matrices satisfy $\{\gamma ^{\mu },\gamma ^{\nu
}\}=2g^{\mu \nu }\times 1$. \ We choose a representation for them in the
form:

\begin{eqnarray}
\gamma ^{t} &=&\frac{1}{\sqrt{F(r,\theta )}}\gamma ^{0}\text{ \ \ }\gamma
^{r}=\sqrt{g(r,\theta )}\gamma ^{3}\text{ \ \ \ \ }\gamma ^{\theta }=\frac{1%
}{\sqrt{\Sigma (r,\theta )}}\gamma ^{1}  \notag \\
\gamma ^{\phi } &=&\frac{1}{\sqrt{K(r,\theta )}}\left( \gamma ^{2}+\frac{%
H(r,\theta )}{\sqrt{F(r,\theta )K(r,\theta )}}\gamma ^{0}\right)
\label{curvedgammas}
\end{eqnarray}%
where the $\gamma ^{a}$'s are simply the following chiral $\gamma $'s for
Minkowski space

\begin{eqnarray}
\gamma ^{0} &=&\left( 
\begin{array}{cc}
0 & I \\ 
-I & 0%
\end{array}%
\right) \text{ \ \ \ }\gamma ^{1}=\left( 
\begin{array}{cc}
0 & \sigma ^{1} \\ 
\sigma ^{1} & 0%
\end{array}%
\right)  \notag \\
\gamma ^{2} &=&\left( 
\begin{array}{cc}
0 & \sigma ^{2} \\ 
\sigma ^{2} & 0%
\end{array}%
\right) \text{ \ }\gamma ^{3}=\left( 
\begin{array}{cc}
0 & \sigma ^{3} \\ 
\sigma ^{3} & 0%
\end{array}%
\right)  \label{chiralgammas}
\end{eqnarray}%
and the $\sigma ^{\prime }s$ are the Pauli Matrices 
\begin{equation}
\sigma ^{1}=\left( 
\begin{array}{cc}
0 & 1 \\ 
1 & 0%
\end{array}%
\right) \text{ \ \ \ \ }\sigma ^{2}=\left( 
\begin{array}{cc}
0 & -i \\ 
i & 0%
\end{array}%
\right) \text{ \ \ \ }\sigma ^{3}=\left( 
\begin{array}{cc}
1 & 0 \\ 
0 & -1%
\end{array}%
\right)  \label{paulis}
\end{equation}%
and we denote $\xi _{\uparrow /\downarrow }$ for the eigenvectors of $\sigma
^{3}$. Note that\textbf{\ 
\begin{equation}
\gamma ^{5}=i\gamma ^{t}\gamma ^{r}\gamma ^{\theta }\gamma ^{\phi }=\sqrt{%
\frac{g}{FK\Sigma }}\left( 
\begin{array}{cc}
-I+\frac{H}{\sqrt{FK}}\sigma ^{2} & 0 \\ 
0 & I+\frac{H}{\sqrt{FK}}\sigma ^{2}%
\end{array}%
\right)  \label{gam5}
\end{equation}%
}is the resulting $\gamma ^{5}$ matrix.

The spin up (i.e. +ve $r$-direction) ansatz for the Dirac field, has the
form: 
\begin{eqnarray}
\psi _{\uparrow }(t,r,\theta ,\phi ) &=&\left[ 
\begin{array}{c}
A(t,r,\theta ,\phi )\xi _{\uparrow } \\ 
B((t,r,\theta ,\phi )\xi _{\uparrow }%
\end{array}%
\right] \exp \left[ \frac{i}{\hbar }I_{\uparrow }(t,r,\theta ,\phi )\right] 
\notag \\
&=&\left[ 
\begin{array}{c}
A(t,r,\theta ,\phi ) \\ 
0 \\ 
B(t,r,\theta ,\phi ) \\ 
0%
\end{array}%
\right] \exp \left[ \frac{i}{\hbar }I_{\uparrow }(t,r,\theta ,\phi )\right]
\label{spin up}
\end{eqnarray}%
In order to apply the WKB\ approximation we insert the ansatz (\ref{spin up}%
)\ for spin up particles\textbf{\ }into the Dirac Equation. Dividing by the
exponential term and multiplying by $\hbar $ the resulting equations to
leading order in $\hbar $\ are%
\begin{eqnarray}
0 &=&-B\left[ \frac{1}{\sqrt{F(r,\theta )}}\partial _{t}I_{\uparrow }+\sqrt{%
g(r,\theta )}\partial _{r}I_{\uparrow }+\frac{H(r,\theta )}{K(r,\theta )%
\sqrt{F(r,\theta )}}\partial _{\phi }I_{\uparrow }\right.  \notag \\
&&\left. +\frac{qer}{\Sigma (r,\theta )\sqrt{F(r,\theta )}}\left( 1-\frac{%
H(r,\theta )}{K(r,\theta )}a\sin ^{2}(\theta )\right) \right] +Am
\label{spin1} \\
0 &=&-B\left[ \frac{i}{\sqrt{K(r,\theta )}}(\partial _{\phi }I_{\uparrow }-%
\frac{qer}{\Sigma (r,\theta )}a\sin ^{2}\theta )+\frac{1}{\sqrt{\Sigma
(r,\theta )}}\partial _{\theta }I_{\uparrow }\right]  \label{spin2} \\
0 &=&A\left[ \frac{1}{\sqrt{F(r,\theta )}}\partial _{t}I_{\uparrow }-\sqrt{%
g(r,\theta )}\partial _{r}I_{\uparrow }+\frac{H(r,\theta )}{K(r,\theta )%
\sqrt{F(r,\theta )}}\partial _{\phi }I_{\uparrow }\right.  \notag \\
&&\left. +\frac{qer}{\Sigma (r,\theta )\sqrt{F(r,\theta )}}\left( 1-\frac{%
H(r,\theta )}{K(r,\theta )}a\sin ^{2}(\theta )\right) \right] +Bm
\label{spin3} \\
0 &=&-A\left[ \frac{i}{\sqrt{K(r,\theta )}}(\partial _{\phi }I_{\uparrow }-%
\frac{qer}{\Sigma (r,\theta )}a\sin ^{2}\theta )+\frac{1}{\sqrt{\Sigma
(r,\theta )}}\partial _{\theta }I_{\uparrow }\right]  \label{spin4}
\end{eqnarray}%
Note that although $A,B$ are not constant, their derivatives -- and the
components $\Omega _{\mu }$ -- are all of order $O(\hbar )$ and so can be
neglected to lowest order in WKB.

When $m\neq 0$ equations (\ref{spin1}) and (\ref{spin3}) couple whereas when 
$m=0$ they decouple. We employ the ansatz 
\begin{equation}
I_{\uparrow }=-Et+J\phi +W(r,\theta )  \label{spinansatz}
\end{equation}%
and insert it into equations (\ref{spin1}-\ref{spin4}) (where we consider
only the positive frequency contributions without loss of\textbf{\ }%
generality). To simplify the expressions we expand the equations near the
horizon and find 
\begin{eqnarray}
0 &=&-B\left( \frac{(-E+\Omega _{H}J+\frac{qer_{+}}{r_{+}^{2}+a^{2}})}{\sqrt{%
F_{r}(r_{+},\theta )(r-r_{+})}}+\sqrt{g_{r}(r_{+},\theta )(r-r_{+})}%
W_{r}(r,\theta )\right) +Am  \label{spin5} \\
0 &=&-B\left( \frac{i}{\sqrt{K(r_{+},\theta )}}(J-\frac{qer_{+}}{\Sigma
(r_{+},\theta )}a\sin ^{2}\theta )+\frac{1}{\sqrt{\Sigma (r_{+},\theta )}}%
W_{\theta }(r,\theta )\right)  \label{spin6} \\
0 &=&A\left( \frac{(-E+\Omega _{H}J+\frac{qer_{+}}{r_{+}^{2}+a^{2}})}{\sqrt{%
F_{r}(r_{+},\theta )(r-r_{+})}}-\sqrt{g_{r}(r_{+},\theta )(r-r_{+})}%
W_{r}(r,\theta )\right) +Bm  \label{spin7} \\
0 &=&-A\left( \frac{i}{\sqrt{K(r_{+},\theta )}}(J-\frac{qer_{+}}{\Sigma
(r_{+},\theta )}a\sin ^{2}\theta )+\frac{1}{\sqrt{\Sigma (r_{+},\theta )}}%
W_{\theta }(r,\theta )\right)  \label{spin8}
\end{eqnarray}%
where%
\begin{align*}
g_{r}(r_{+},\theta )& =\frac{\Delta _{r}(r_{+})}{\Sigma (r_{+},\theta )}=%
\frac{2r_{+}-2M}{r_{+}^{2}+a^{2}\cos ^{2}(\theta )} \\
F_{r}(r_{+},\theta )& =\frac{\Delta _{r}(r_{+})\Sigma (r_{+},\theta )}{%
(r_{+}^{2}+a^{2})^{2}}=\frac{(2r_{+}-2M)(r_{+}^{2}+a^{2}\cos ^{2}(\theta ))}{%
(r_{+}^{2}+a^{2})^{2}}
\end{align*}

In the massless case it is possible to pull $\frac{1}{\sqrt{\Sigma
(r_{+},\theta )}}$ out of equations (\ref{spin5}) and (\ref{spin7}), making
these equations independent of $\theta $. \ Furthermore, equations (\ref%
{spin6}) and (\ref{spin8}) have no explicit $r$ dependence. \ From this we
can conclude that near the black horizon it is possible to further separate
the function $W$

\begin{equation*}
W(r,\theta )=W(r)+\Theta (\theta )
\end{equation*}%
and we see that equations (\ref{spin6}) and (\ref{spin8}) both yield the
same equation for $\Theta $ regardless of $A$ or $B$.

\begin{figure}[tbp]
\centering\includegraphics[
width=3.9721in
]{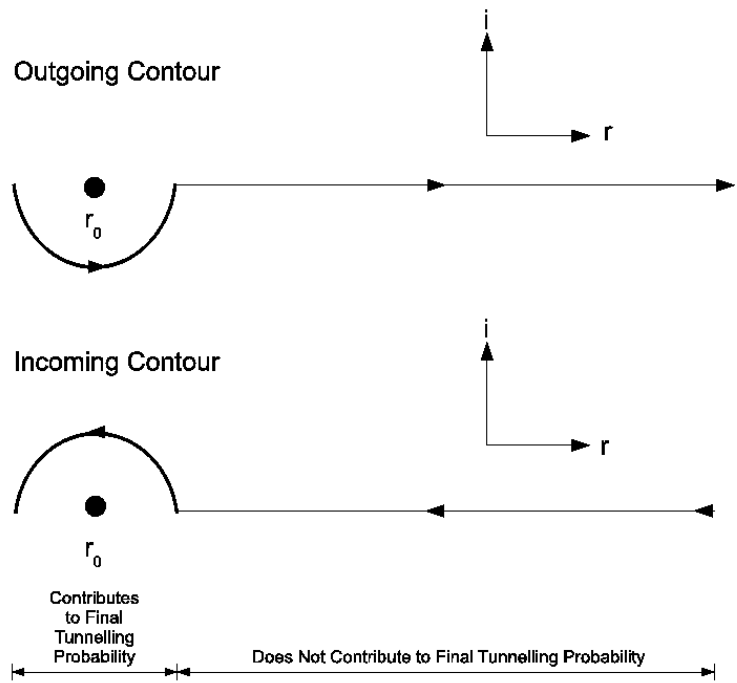}
\caption{Diagram of contours between black hole and observer for both
outgoing and incoming trajectories }
\label{Figure01}
\end{figure}

Equations (\ref{spin5}) and (\ref{spin7}) then have two possible solutions%
\textbf{\ }%
\begin{eqnarray*}
A &=&0\text{ and }W^{\prime }(r)=W_{+}^{\prime }(r)=\frac{(E-\Omega _{H}J-%
\frac{qer_{+}}{r_{+}^{2}+a^{2}})(r_{+}^{2}+a^{2})}{\Delta
_{r}(r_{+})(r-r_{+})} \\
B &=&0\text{ and }W^{\prime }(r)=W_{-}^{\prime }(r)=\frac{-(E-\Omega _{H}J-%
\frac{qer_{+}}{r_{+}^{2}+a^{2}})(r_{+}^{2}+a^{2})}{\Delta
_{r}(r_{+})(r-r_{+})}
\end{eqnarray*}%
where the prime denotes a derivative with respect to $r$ and $W_{+/-}$
corresponds to outgoing/incoming solutions. Since we are only concerned with
calculating the semi-classical tunnelling probability, we will need to
multiply the resulting wave equation by its complex conjugate. So the
portion of the trajectory that starts outside the black hole and continues
to the observer will not contribute to the final tunnelling probability and
can be safely ignored (since it will be entirely real). \ Therefore, the
only part of the wave equation that contributes to the tunnelling
probability is the contour around the black hole horizon. \ For a visual
representation of the deformation of the contour see Figure \ref{Figure01}.
\ This contour differs somewhat relative to conventions Padmanabhan defines 
\cite{Padmanabhan}, in which contours in the upper half plane are selected
for both ingoing and outgoing particles. For the emission integral he
multiplies his equivalent of $W_{+}^{^{\prime }}$ ($\partial S_{0}/\partial
r $) by a minus sign\textquotedblleft where the minus sign in front of the
integral corresponds to the initial condition that $\partial S_{0}/\partial
r>0$\ at\textbf{\ }$r=r_{1}<r_{0}$" \cite{Padmanabhan}. Instead we choose
the mathematically equivalent convention that the outgoing contour is in the
lower half plane and so do not multiply by a minus sign. 

The probabilities of crossing the horizon in each direction are proportional
to 
\begin{eqnarray}
\text{Prob}[out] &\propto &\exp [-2\func{Im}I]=\exp [-2(\func{Im}W_{+}+\func{%
Im}\Theta )]  \label{outprob} \\
\text{Prob}[in] &\propto &\exp [-2\func{Im}I]=\exp [-2(\func{Im}W_{-}+\func{%
Im}\Theta )]  \label{inprob}
\end{eqnarray}

To ensure that the probabilities are correctly normalized so that any
incoming particles crossing the horizon have a $100\%$ chance of entering
the black hole we need to divide each equation by (\ref{inprob}). \ From
this the probability of going from outside to inside the horizon will be
equal to $1$ and \ this implies that the probability of a particle
tunnelling from inside to outside the horizon is:%
\begin{equation}
\Gamma \propto \frac{\text{Prob}[out]}{\text{Prob}[in]}=\frac{\exp [-2(\func{%
Im}W_{+}+\func{Im}\Theta )]}{\exp [-2(\func{Im}W_{-}+\func{Im}\Theta )]}%
=\exp [-4\func{Im}W_{+}]
\end{equation}%
Solving for $W_{+}$ yields 
\begin{equation*}
W_{+}(r)=\int \frac{(E-\Omega _{H}J-\frac{qer_{+}}{r_{+}^{2}+a^{2}}%
)(r_{+}^{2}+a^{2})}{\Delta _{r}(r_{+})(r-r_{+})}
\end{equation*}%
and after integrating around the pole (and dropping the + subscript) we
obtain 
\begin{eqnarray}
W &=&\frac{\pi i(E-\Omega _{H}J-\frac{qer_{+}}{r_{+}^{2}+a^{2}}%
)(r_{+}^{2}+a^{2})}{2r_{+}-2M}  \notag \\
\func{Im}W &=&(E-\Omega _{H}J-\frac{qer_{+}}{r_{+}^{2}+a^{2}})\frac{\pi }{2}%
\frac{r_{+}^{2}+a^{2}}{(r_{+}-M)}  \label{imW}
\end{eqnarray}

The resulting tunnelling probability is 
\begin{equation*}
\Gamma =\exp [-2\pi \frac{r_{+}^{2}+a^{2}}{(r_{+}-M)}(E-\Omega _{H}J-\frac{%
qer_{+}}{r_{+}^{2}+a^{2}})]
\end{equation*}%
giving the expected Hawking temperature 
\begin{equation}
T_{H}=\frac{1}{2\pi }\frac{r_{+}-M}{r_{+}^{2}+a^{2}}=\frac{1}{2\pi }\frac{%
(M^{2}-a^{2}-e^{2})^{\frac{1}{2}}}{2M(M+(M^{2}-a^{2}-e^{2})^{\frac{1}{2}%
})-e^{2}}  \label{temp}
\end{equation}%
for a charged rotating black hole.

In the massive case equations (\ref{spin5}) and (\ref{spin7}) no longer
decouple and analysis of the tunnelling is more subtle. \ We begin by
eliminating the function $W_{r}(r,\theta )$ from these two equations so we
can find an equation relating $A$ and $B$ in terms of known quantities.
Subtracting $B\times $(\ref{spin7}) from $A\times $\ (\ref{spin5}) gives 
\begin{eqnarray}
0 &=&\frac{2AB(E-\Omega _{H}J-\frac{qer_{+}}{r_{+}^{2}+a^{2}})}{\sqrt{%
F_{r}(r_{+},\theta )(r-r_{+})}}+mA^{2}-mB^{2}=0  \label{ab1} \\
0 &=&m\sqrt{F_{r}(r_{+},\theta )(r-r_{+})}(\frac{A}{B})^{2}+2(E-\Omega _{H}J-%
\frac{qer_{+}}{r_{+}^{2}+a^{2}})(\frac{A}{B})  \notag \\
&&-m\sqrt{F_{r}(r_{+},\theta )(r-r_{+})}  \label{AB2}
\end{eqnarray}%
and so 
\begin{equation*}
\frac{A}{B}=\frac{-(E-\Omega _{H}J-\frac{qer_{+}}{r_{+}^{2}+a^{2}})\pm \sqrt{%
(E-\Omega _{H}J-\frac{qer_{+}}{r_{+}^{2}+a^{2}})^{2}+m^{2}F_{r}(r_{+},\theta
)(r-r_{+})}}{m\sqrt{F_{r}(r_{+},\theta )(r-r_{+})}}
\end{equation*}%
where 
\begin{equation}
\lim_{r\rightarrow r_{+}}\left( \frac{-(E-\Omega _{H}J-\frac{qer_{+}}{%
r_{+}^{2}+a^{2}})\pm \sqrt{(E-\Omega _{H}J-\frac{qer_{+}}{r_{+}^{2}+a^{2}}%
)^{2}+m^{2}F_{r}(r_{+},\theta )(r-r_{+})}}{m\sqrt{F_{r}(r_{+},\theta
)(r-r_{+})}}\right) =\left\{ 
\begin{array}{c}
0 \\ 
-\infty%
\end{array}%
\right.  \label{limit}
\end{equation}%
for the upper/lower sign respectively.

Consequently at the horizon either $\frac{A}{B}\rightarrow 0$ or $\frac{A}{B}%
\rightarrow -\infty $, i.e. either $A\rightarrow 0$ or $B\rightarrow 0$. For 
$A\rightarrow 0$ at the horizon, we solve (\ref{spin7}) in terms of $m$ and
insert into (\ref{spin5}), obtaining 
\begin{equation}
W_{r}(r,\theta )=\frac{(E-\Omega _{H}J-\frac{qer_{+}}{r_{+}^{2}+a^{2}})}{%
\sqrt{F_{r}(r_{+},\theta )g_{r}(r_{+},\theta )}(r-r_{+})}\frac{\left( 1+%
\frac{A^{2}}{B^{2}}\right) }{\left( 1-\frac{A^{2}}{B^{2}}\right) }
\label{W massive}
\end{equation}%
Note that the $\theta $-dependence drops out of this expression, i.e.%
\begin{equation*}
W_{r}(r,\theta )\equiv {W}_{+}^{\prime }(r)=\frac{(E-\Omega _{H}J-\frac{%
qer_{+}}{r_{+}^{2}+a^{2}})}{\sqrt{F_{r}(r_{+},\theta )g_{r}(r_{+},\theta )}%
(r-r_{+})}\frac{\left( 1+\frac{A^{2}}{B^{2}}\right) }{\left( 1-\frac{A^{2}}{%
B^{2}}\right) }
\end{equation*}%
since $\frac{A}{B}$ is zero at the horizon the result of integrating around
the pole is the same as in the massless case. \ For $B\rightarrow 0$ we can
simply rewrite the expression (\ref{W massive}) in terms of $\frac{B}{A}$ to
get

\begin{equation*}
W_{r}(r,\theta )\equiv {W}_{-}^{\prime }(r)=\frac{-(E-\Omega _{H}J-\frac{%
qer_{+}}{r_{+}^{2}+a^{2}})}{\sqrt{F_{r}(r_{+},\theta )g_{r}(r_{+},\theta )}%
(r-r_{+})}\frac{\left( 1+\frac{B^{2}}{A^{2}}\right) }{\left( 1-\frac{B^{2}}{%
A^{2}}\right) }
\end{equation*}%
Again, since the extra contributions vanish at the horizon, the result of
integrating around the pole for $W$ in the massive case is the same as the
massless case and we recover the Hawking temperature (\ref{temp}) for the
Kerr-Newman black hole. \ 

The spin-down case proceeds in a manner fully analogous to the spin-up case
discussed above. Other than some changes of sign the equations are of the
same form as the spin up case. For both the massive and massless cases the
temperature (\ref{temp}) is obtained, implying that both spin up and spin
down particles are emitted at the same rate.

\section{Technical Issues}

With rotating spacetimes the choice $\gamma $ matrices is quite relevant,
not only for ease of calculation but also for tractability. In order to
demonstrate this we will repeat the calculation for a different (and less
convenient) choice of $\gamma $ matrices. \ We will also set the charge $q$\
of the emitted particles to zero for simplicity.

Consider the choice 
\begin{eqnarray}
\tilde{\gamma}^{t} &=&\frac{1}{\sqrt{f(r,\theta )}}\left( \gamma ^{0}-\frac{%
H(r,\theta )}{\sqrt{F(r,\theta )K(r,\theta )}}\gamma ^{2}\right) \text{ \ \
\ \ \ }\tilde{\gamma}^{r}=\sqrt{g(r,\theta )}\gamma ^{3}  \notag \\
\tilde{\gamma}^{\theta } &=&\frac{1}{\sqrt{\Sigma (r,\theta )}}\gamma ^{1}%
\text{ \ \ \ \ \ \ \ \ \ \ \ \ \ \ \ \ \ \ \ \ \ \ \ \ \ \ \ \ \ \ \ }\tilde{%
\gamma}^{\phi }=\sqrt{\frac{f(r,\theta )}{F(r,\theta )K(r,\theta )}}\gamma
^{2}  \label{gamchoice2}
\end{eqnarray}%
where we use the same $\ $chiral $\gamma ^{a}$ matrices (\ref{chiralgammas})
as before. This choice satisfies the correct anti-commutation relations $\{%
\tilde{\gamma}^{\mu },\tilde{\gamma}^{\nu }\}=2g^{\mu \nu }$, and
corresponds to a different choice of tetrad basis for the metric. \ 

Naively choosing the same ansatz as before

\begin{equation*}
\psi _{\uparrow }(t,r,\theta ,\phi )=\left[ 
\begin{array}{c}
A \\ 
0 \\ 
B \\ 
0%
\end{array}%
\right] \exp \left[ \frac{i}{\hbar }I\right]
\end{equation*}%
we find upon insertion into the (chargeless) Dirac equation (\ref{Dirac}) we
obtain

\begin{eqnarray}
-B\left( \frac{1}{\sqrt{f}}\partial _{t}I+\sqrt{g}\partial _{r}I\right) +Am
&=&0  \label{flawed1} \\
-B\left( i\sqrt{\frac{f}{FK}}\partial _{\phi }I-i\frac{H}{\sqrt{FK}}\partial
_{t}I+\frac{1}{\sqrt{\Sigma }}\partial _{\theta }I\right) &=&0 \\
A\left( \frac{1}{\sqrt{f}}\partial _{t}I-\sqrt{g}\partial _{r}I\right) +Bm
&=&0  \label{flawed2} \\
-A\left( i\sqrt{\frac{f}{FK}}\partial _{\phi }I-i\frac{H}{\sqrt{FK}}\partial
_{t}I+\frac{1}{\sqrt{\Sigma }}\partial _{\theta }I\right) &=&0
\end{eqnarray}
Repeating the same kind of analysis as before (using the ansatz $I=-Et+J\phi
+W$ ) we find from (\ref{flawed1}) and (\ref{flawed2}) that%
\begin{equation*}
W_{r\pm }=\frac{\pm E}{\sqrt{fg}}
\end{equation*}%
Since $f$ does not vanish at the horizon (except when $\sin \theta =0)$,
this expression does not have a simple pole at the horizon. It is not
possible to solve the expression for arbitrary $\theta $, and the
calculation becomes intractable. This situation is analogous to what happens
in the scalar field case if one naively applies the null geodesic method to
a rotating black hole by trying to force $\phi $ to be constant (id $d\phi
=0 $), as previously demonstrated \cite{first paper}. \ 

In order to understand this issue in more detail it is useful to examine the
similarity transformation between $\gamma ^{\mu }$ and $\tilde{\gamma}^{\mu
} $. \ We find that:

\begin{equation*}
\tilde{\gamma}^{\mu }=S\gamma ^{\mu }S^{-1},\text{ for all }\mu
\end{equation*}

when:

\begin{equation*}
S=\left( 
\begin{array}{cc}
aI-b\sigma ^{2} & 0 \\ 
0 & aI+b\sigma ^{2}%
\end{array}%
\right)
\end{equation*}%
where 
\begin{equation}
a=\sqrt{\frac{1}{2}\left( \sqrt{\frac{F}{f}}+1\right) }\text{ \ \ \ \ \ }b=%
\sqrt{\frac{1}{2}\left( \sqrt{\frac{F}{f}}-1\right) }
\end{equation}

The transformation $S$ is similar to a Lorentz boost in the $\phi $
direction. Applying it to the spin up ansatz used previously we find

\begin{equation}
\tilde{\psi}_{\uparrow }(t,r,\theta ,\phi )=\left[ 
\begin{array}{c}
Aa \\ 
-Aib \\ 
Ba \\ 
Bib%
\end{array}%
\right] \exp \left[ \frac{i}{\hbar }I\right]  \label{newansatz}
\end{equation}%
As $r\rightarrow \infty $ we see that $a\rightarrow 1$ and $b\rightarrow 0$,
yielding the same spin up ansatz in this limit. \ Inserting (\ref{newansatz}%
) into the (chargeless) Dirac equation (\ref{Dirac}) and following the same
procedure as before results in the same expression (\ref{temp})\ for the
temperature. \ This is not surprising since all we have done is applied a
similarity transformation to the Dirac equation, and we shall not repeat the
(somewhat more tedious) calculations here. \ Our point is to emphasize the
importance of choosing an appropriate ansatz for a given choice of $\gamma $
matrices. \ \ \ \ 

\section{Conclusions}

We have successfully extended our approach of fermion tunnelling to model
the emission of charged fermions from a rotating charged black hole. The
analysis yields the expected temperature (\ref{temp}), consistent with black
hole universality. However there are subtle technical issues involved with
choosing an appropriate ansatz for the Dirac field consistent with the
choice of \ $\gamma $\ matrices, and failure to make such a choice leads to
a breakdown in the method.

Computing corrections to the tunnelling probability by fully taking into
account conservation of energy will yield corrections to the fermion
emission temperature. \ In various scalar field cases this is inherent in
the Parikh/Wilczek tunnelling method \cite{Parikh}, \cite{Vagenas2}-\cite%
{charged} and can be incorporated into the Hamilton-Jacobi tunnelling
approach \cite{Vagenas1}. \ Another avenue of research is to perform
tunnelling calculations to higher order in WKB (in both the scalar field and
fermionic cases) in order to calculate grey body effects. It is also worth
investigating the possibility of calculating a density matrix for the
emitted particles from a tunnelling approach in order to calculate
correlations between particles. \ Work is in progress in these areas.

{\Huge \bigskip }

{\Huge \medskip Acknowledgements}

This work was supported in part by the Natural Sciences and Engineering
Research Council of Canada. As this work was being complete we became aware
of similar work on fermion emission from Kerr-Newman black holes \cite{ferm
kerr} in which results similar to ours were obtained.\ {\Huge \bigskip }


\begin{thebibliography}{99}
\bibitem{early tunnelling} P. Kraus and F. Wilczek,\textquotedblleft A
Simple Stationary Line Element for the Schwarzschild Geometry, and Some
Applications\textquotedblright\ [gr-qc/9406042]; P. Kraus and F. Wilczek,
\textquotedblleft Self-Interaction Correction to Black Hole
Radiance\textquotedblright , Nucl. Phys. \textbf{B433}, 403 (1995)
[gr-qc/9408003]; P. Kraus and F. Wilczek, \textquotedblleft Effect of
Self-Interaction on Charged Black Hole Radiance\textquotedblright , Nucl.
Phys. \textbf{B437 } 231-242 (1995) [hep-th/9411219]; P. Kraus and E.
Keski-Vakkuri, \textquotedblleft Microcanonical D-branes and Back
Reaction\textquotedblright , Nucl.Phys. \textbf{B491} 249-262 (1997)
[hep-th/9610045]

\bibitem{late 90s tunnelling} V. A. Berezin, A. Boyarsky, and A. Yu.
Neronov, "On the Mechanism of Hawking Radiation", Gravitation \& Cosmology,
Vol. 5 16-22 (1999); G. E. Volovik, "Simulation of Panleve-Gullstrand black
hole in thin 3He-A film", Pisma Zh.Eksp.Teor.Fiz. 69 (1999) 662-668; JETP
Lett. 69 (1999) 705-713; \ A. Calogeracos, and G.E. Volovik, "Rotational
quantum friction in superfluids: Radiation from object rotating in
superfluid vacuum", JETP Lett. 69 (1999) 281-287; Pisma Zh.Eksp.Teor.Fiz. 69
(1999) 257-262\ 

\bibitem{Parikh} M. K. Parikh and F. Wilczek, \textquotedblleft Hawking
Radiation as Tunneling\textquotedblright , Phys. Rev. Lett. \textbf{85},
5042 (2000), [arXiv:hep-th/9907001];M. K. Parikh, \textquotedblleft New
Coordinates for de Sitter Space and de Sitter Radiation\textquotedblright ,
Phys. Lett. B \textbf{546,} 189, (2002) [hep-th/0204107]; M. K. Parikh,
\textquotedblleft A Secret Tunnel Through The Horizon\textquotedblright ,
Int.J.Mod.Phys. \textbf{D13} 2351-2354 (2004) [hep-th/0405160]; M. K.
Parikh, \textquotedblleft Energy Conservation and Hawking
Radiation\textquotedblright , [arXiv:hep-th/0402166];

\bibitem{Padmanabhan} K. Srinivasan and T.Padmanabhan , \textquotedblleft
Particle Production and Complex Path Analysis\textquotedblright , Phys. Rev. 
\textbf{D60} , 24007 (1999) [gr-qc-9812028]; S. Shankaranarayanan, K.
Srinivasan, and T. Padmanabhan, "Method of complex paths and general
covariance of Hawking radiation" \ Mod.Phys.Lett. A16 (2001) 571-578
[arXiv:gr-qc/0007022v2]; S. Shankaranarayanan, T. Padmanabhan, and K.
Srinivasan, "Hawking radiation in different coordinate settings: Complex
paths approach" Class.Quant.Grav. 19 (2002) 2671-2688
[arXiv:gr-qc/0010042v4]; S. Shankaranarayanan, "Temperature and entropy of
Schwarzschild-de Sitter space-time" Phys.Rev. D67 (2003) 084026
[arXiv:gr-qc/0301090v2]

\bibitem{parikh1} M. K. Parikh, \textquotedblleft New Coordinates for de
Sitter Space and de Sitter Radiation\textquotedblright , Phys. Lett. B 
\textbf{546,} 189, (2002) [arXiv:hep-th/0204107]

\bibitem{schwdS} A.J.M. Medved, "Radiation via tunnelling from a de Sitter
cosmological horizon.", \ Phys. Rev. \textbf{D66}: 124009 (2002) [arXiv:
hep-th/0207247

\bibitem{Vanzo} M. Agheben, M. Nadalini, L Vanzo, and S. Zerbini,
\textquotedblleft Hawking Radiation as Tunneling for Extremal and Rotating
Black Holes\textquotedblright , JHEP 0505 (2005) 014 [hep-th/0503081];

\bibitem{Vagenas1} A.J.M. Medved and E.Vagenas, "On Hawking radiation as
tunneling with back-reaction", Mod. Phys. Lett. A20:2449-2454, (2005)

\bibitem{Vagenas2} M. Arzano, A. Medved and E. Vagenas, \ \textquotedblleft
Hawking Radiation as Tunneling through the Quantum Horizon\textquotedblright
, JHEP 0509 (2005) 037 [hep-th/0505266]

\bibitem{kerr and kerr newman} Qing-Quan Jiang, Shuang-Qing Wu, and Xu Cai,
``Hawking radiation as tunneling from the Kerr and Kerr-Newman black
holes'',Phys.Rev. D73 (2006) 064003 \ [hep-th/0512351]

\bibitem{Zhang and Zhao} Jingyi Zhang, and Zheng Zhao, \textquotedblleft
Charged particles' tunnelling from the Kerr-Newman black
hole\textquotedblright , Phys.Lett. B638 (2006) 110-113 [gr-qc/0512153]; \ \
Yapeng Hu, Jingyi Zhang, and Zheng Zhao, "The relation between Hawking
radiation via tunnelling and the laws of black hole thermodynamics"
[gr-qc/0601018]

\bibitem{Black Rings} Liu Zhao, "Tunnelling through black rings",
[hep-th/0602065]

\bibitem{BTZ} Shuang-Qing Wu, and Qing-Quan Jiang, "Remarks on Hawking
radiation as tunneling from the BTZ black holes", JHEP 0603 (2006) 079,
[hep-th/0602033]

\bibitem{first paper} R. Kerner and R.B. Mann, \textquotedblleft Tunnelling,
Temperature and Taub-NUT Black Holes\textquotedblright , \ Phys.Rev. \textbf{%
D73} (2006) 104010

\bibitem{Higher D R-N} Shuang-Qing Wu, and Qing-Quan Jiang, "Hawking
Radiation of Charged Particles as Tunneling from Higher Dimensional
Reissner-Nordstrom-de Sitter Black Holes", \ [hep-th/0603082]

\bibitem{issues with tunnelling method} B. D. Chowdhury, "Tunneling of Thin
Shells from Black Holes: An Ill Defined Problem", [hep-th/0605197]

\bibitem{rotating} Satoshi Iso, Hiroshi Umetsu, and Frank Wilczek,
"Anomalies, Hawking Radiations and Regularity in Rotating Black
Holes",Phys.Rev. \textbf{D74} (2006) 044017 [hep-th/0606018]

\bibitem{Vaidya} Jun Ren, Jingyi Zhang, and Zheng Zhao, "Tunnelling Effect
and Hawking Radiation from a Vaidya Black Hole", Chin.Phys.Lett. 23 (2006)
2019-2022, [gr-qc/0606066]

\bibitem{dynamicalbh} R. Di Criscienzo, M. Nadalini, L. Vanzo, S. Zerbini,
and G. Zoccatelli, "On the Hawking radiation as tunneling for a class of
dynamical black holes", \ [arXiv: 0707.4425 [hep-th]]

\bibitem{charged} Zhao Ren, Li Huai-Fan, and Zhang Sheng-Li, "Canonical
Entropy of charged black hole" [gr-qc/0608123]

\bibitem{Mitra} P. Mitra, ''Hawking temperature from tunnelling formalism'',
Phys. Lett. \textbf{B648}:240-242, (2007) [hep-th/0611265]; Bhramar
Chatterjee, Amit Ghosh, P. Mitra, `` Tunnelling from black holes in the
Hamilton Jacobi approach'', arXiv:0704.1746 [hep-th]

\bibitem{Godel} R. Kerner and R.B. Mann, \textquotedblleft Tunnelling From G%
\"{o}del Black Holes\textquotedblright , \ Phys. Rev. \textbf{D75}: 084022
,(2007)

\bibitem{Majhi} R.Banerjee and B.R.Majhi \textquotedblleft Quantum Tunneling
and Back Reaction\textquotedblright , to appear in Phys. Lett. B,
arXiv:0801.0200; R.Banerjee, B.R.Majhi and S.Samanta, \textquotedblleft
Noncommutative Black Hole

Thermodynamics\textquotedblright , arXiv:0801.3583

\bibitem{Fermion tunnelling} Ryan Kerner, R.B. Mann, ``Fermions Tunnelling
from Black Holes'', \ arXiv:0710.0612v4

\bibitem{ferm cosmo hor} Yuichi Sekiwa, ``Decay of the cosmological constant
by Hawking radiation as quantum tunneling'', arXiv:0802.3266v1

\bibitem{ferm BTZ} Ran Li and Ji-Rong Ren, ``Dirac particles tunneling from
BTZ black hole'', \ \qquad arXiv:0802.3954v1

\bibitem{ferm variable mass} Roberto Di Criscienzo and Luciano Vanzo,
\textquotedblleft Fermion Tunneling from Dynamical
Horizons\textquotedblright , arXiv:0803.0435

\bibitem{ferm kerr} Ran Li and Ji-Rong Ren, \textquotedblleft Hawking
radiation of Dirac particles via tunneling from Kerr black
hole\textquotedblright , arXiv:0803.1410

\bibitem{hawking} S. W. Hawking, Commun., \textquotedblleft Particle
Creation By Black Holes\textquotedblright , Math. Phys. \textbf{43}, 199
(1975)

\bibitem{Unruh} \ W. G. Unruh, \textquotedblleft Notes on black hole
evaporation\textquotedblright , Phys. Rev. \textbf{D14}, 870 (1976)

\bibitem{alsingpaper} P. Alsing, I.\ Fuentes-Schuller, R.B. Mann and T.
Tessier, \textquotedblleft Entanglement of Dirac fields in non-inertial
frames\textquotedblright , Phys. Rev. \textbf{A74} (2006) 032326.
\end{thebibliography}
\end{document}